\documentclass[floatfix,prl,aps,twocolumn]{revtex4}
\usepackage{graphicx}
\usepackage{psfrag}

\begin{document}

\title{Fast quantum limited read-out of a superconducting qubit
 using a slow oscillator}

\author{G{\"o}ran Johansson, Lars Tornberg, C.M. Wilson}

\affiliation{Microtechnology and Nanoscience, MC2, Chalmers, S-412 96
G\"oteborg, Sweden}

\begin{abstract}
We describe how to perform fast quantum limited read-out of a
solid state qubit biased at its degeneracy point. The method is
based on homodyne detection of the phase of a microwave signal
reflected by a slow oscillator coupled to the qubit. Analyzing the
whole quantum read-out process, we find that the detection is
indeed quantum limited and that this limit may be reached even
using a resonance circuit with a low quality factor, thus enabling
the use of short measurement pulses. As an example, we discuss in
detail the read-out of a Cooper-pair box capacitively coupled to a
lumped element LC-oscillator.
%Here the read-out is equivalent to measuring the box's quantum capacitance, as shown by
%recent experiments (T. Duty et al. [Phys. Rev. Lett. 95, 206807 (2005)] and
%M. Sillanp{\"a}{\"a} et al. [Phys. Rev. Lett. 95, 206806 (2005)]).
Furthermore, we give formulas for the backaction while not
measuring, and discuss optimal parameters for a realistic design
capable of fast ($\sim 50$~ns) single-shot read-out.
\end{abstract}

\maketitle

%\section{Introduction}

Superconducting qubits are strong contenders in the race to build
a quantum computer\cite{WendinShumeiko2005}. Accordingly, a great
deal of effort has gone into developing fast and accurate
single-shot read-outs for these devices.
%For an efficient implementation, fast and accurate single-shot
%read out is very important.

Speed is one important characteristic for a read-out. For accurate
qubit detection, the read-out has to be faster than the qubit
relaxation time. Moreover, to implement quantum error correction,
the read-out must be quicker than the decoherence time.

Another figure of merit, which characterizes the
backaction, is the quantum efficiency, $\eta$. According to
fundamental principles of quantum measurement, the measurement
time, $t_{ms}$, needed to distinguish the two qubit states is
always longer or equal to half the dephasing time, $t_\varphi$,
after which the qubit has lost its quantum
coherence\cite{Braginsky}. Thus, the quantum efficiency has an
upper bound of unity, $\eta=t_\varphi/2t_{ms} \leq 1$.

Recently, there has been increasing interest in dispersive
read-outs for superconducting qubits \cite{GrajcarPRB04,
LupascuPRL04, WallraffPRL05, SiddiqiPRB06}, which measure the
reactive response of an oscillator coupled to the qubit. The
energy needed for detection is then dissipated far away from the
qubit, giving the schemes low backaction. Also, many dispersive
read-outs work with the qubit biased at its degeneracy point where
the decoherence induced by low frequency fluctuations is minimized
\cite{NECnoise,SaclayNoise}.

In this letter, we consider how to optimize a dispersive qubit
read-out for speed, while maintaining high quantum effieciency and
minimizing decoherence.  At Yale, Wallraff {\it et al.} coupled a
charge qubit to a cavity made from a coplanar waveguide
(CPW)\cite{WallraffPRL05}. In this experiment, a cavity with a
high quality factor ($Q$) was used to enhance the dispersive phase
shift of detected photons.  However, using a high-$Q$ cavity
severely limits the read-out speed, naturally leading us to
consider using a low-$Q$ resonator instead.  That said, low-$Q$
reduces the dispersive phase shift  and implies the need for large
detuning between the cavity and qubit, because qubit relaxation is
strongly enhanced if the qubit frequency lies within the cavity
resonance. Bertet {\it et al.}\cite{DelftFluxPhotonDephasing}
coupled a flux qubit to a low-frequency SQUID oscillator.
Unfortunately, they found that the thermal photon noise of the
oscillator, present even when not measuring, limited the qubit
coherence.

We attempt to harmonize these apparently contradictory
requirements. The standard methods of cavity QED
\cite{YalePRA,ZollerGardiner} are not appropriate for this
analysis because they only treat the high-$Q$ limit.  Therefore,
we take a new approach, based on the quantum network theory
introduced by Yurke and Denker \cite{YurkeDenkerPRA}, which gives
analytic results for arbitrary values of $Q$. Indeed, we find that
the read-out is quantum limited ($\eta=1$) independent of $Q$. We
go on to calculate the thermal dephasing time in the absence of
measurement, $t^{\rm off}_\varphi$. We find that low $Q$ can
compensate the effects of thermal photon noise on $t^{\rm
off}_\varphi$ by suppressing low-frequency photon fluctuations. We
also find that large detuning compensates for a the smaller phase
shift by accommodating stronger driving. We go on to discuss
optimal parameters for a realistic single-shot read-out.

Our approach is general, and we use it to derive both the
Hamiltonians for the measurement of the effective capacitance
(inductance) of a charge (flux) qubit. However, we focus the
optimization on the Cooper-pair box operated as a charge
qubit\cite{CPB_qubit_ref}, capacitively coupled to a
low-frequency, lumped-circuit LC-resonator.
In this setup, the effective capacitance of the Cooper-pair box
was recently measured by two different groups \cite{DutyPRL,
Sillanpaa}.
\begin{figure}[!ht]
\includegraphics[width=8cm]{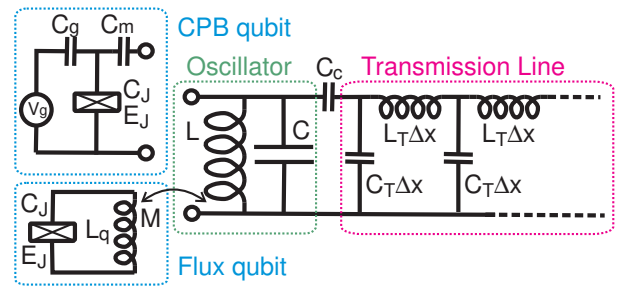}
\caption{A CPB (flux) qubit capacitively (inductively) coupled to an
LC-oscillator, which is attached to a transmission line.}\label{CircuitFig}
\end{figure}

The Cooper-pair box qubit consists of a small superconducting
island coupled to a superconducting reservoir through a Josephson
junction characterized by its capacitance, $C_J$, and Josephson
energy, $E_J$. The island is coupled to a gate voltage, $V_g$,
through the gate capacitance, $C_g$, and to the LC-oscillator via
the coupling capacitance, $C_m$. The oscillator is in turn
coupled, via $C_c$, to a transmission line, here modelled as a
semi-infinite line of LC-oscillators. The transmission line is
characterized by its capacitance, $C_T$, and inductance, $L_T$,
per unit length. (See Fig.~\ref{CircuitFig}.)

%\subsection{Hamiltonian}
Since we have a Josephson junction in our circuit, we choose as
our coordinates the phase differences across the circuit elements
$\phi_\alpha(t) = \int^t dt' V_\alpha(t')$. In the transmission
line, we number the phases across the capacitances starting from
the oscillator, giving the classical
Lagrangian\cite{JohanssonJPCM06}
\begin{eqnarray}
L &=& \frac{C_{qb} \dot{\phi}^2_J}{2} -
\frac{C_m}{2}\dot{\phi}_C\dot{\phi}_J - C_gV_g\dot{\phi}_J + E_J
\cos\phi_J + \nonumber \\
&+& \frac{(C_{osc}+C_c)\dot{\phi}_C^2}{2} + \frac{C_c
\dot{\phi}_1^2}{2} -
\frac{C_c}{2}\dot{\phi}_C\dot{\phi}_1 - \frac{\phi^2_C}{2 L} +\nonumber\\
 &+& \sum_{i=1}^\infty \Delta x \left( \frac{C_T\dot{\phi}_i^2}{2}
 - \frac{(\phi_{i+1} - \phi_{i} )^2}{2L_T(\Delta x)^2}\right),
\end{eqnarray}
where $C_{qb} = C_J + C_g + C_m$, $C_{osc} = C + C_m$.
% and we have used Kirchoff's circuit rules as constraints.
%Through a Legendre transformation, we get the Hamiltonian for the
%circuit, including the transmission line.
Conjugate to each coordinate
$\phi_\alpha$, the momentum $q_\alpha$
has the dimension of charge, and $[\phi_\alpha,q_\beta] = i\hbar
\delta_{\alpha\beta}$.
%To describe quantum dynamics,
%the phase and charge are considered operators that obey, by
%definition, the canonical commutation relation
%$[\phi_\alpha,q_\beta] = i\hbar \delta_{\alpha\beta}$.
%In the Hamiltonian we have the
%following effective electrostatic energies
%\begin{equation}
%E_C=\frac{e^2 C_{osc}}{2(C_{qb}C_{osc}-C_m^2)} \approx
%\frac{e^2}{2C_{qb}}
%\end{equation}

%To realize a charge qubit, we consider the Josephson energy $E_J$
%small compared to the charging energy $E_C = e^2/2C_{qb}$ and
%limit ourselves to two relevant charge states, $|n=0\rangle$ and
%$|n=1\rangle$, with zero and one extra Cooper-pairs in the CPB.
%The qubit part of the Hamiltonian then reads
%\begin{equation}\label{eq:Hq3}
%H_{q} =  -2E_C\left(1 - \frac{C_g V_g}{e} - \kappa
%\frac{q_C+q_1}{e}\right)\sigma_z - \frac{E_J}{2}\sigma_x,
%\end{equation}
%where $\kappa=C_m/C_{osc}\ll 1$ and $\sigma_x$ and $\sigma_z$ are
%Pauli spin operators in the charge basis.
During read-out the oscillator and transmission line charges
($q_C$ and $q_1$) will be driven harmonically at a frequency close
to the bare oscillator frequency $\omega_0 =
1/\sqrt{L(C_{osc}+C_c)}$, which we consider much lower than the
qubit frequency $\omega_{qb}=E_J/\hbar$. The amplitude of
 the charge oscillations induced on the qubit island is determined
 by the driving strength, which we characterize by a parameter $\beta$
 through
\begin{equation}
\label{drive_strength_limit} \kappa \frac{\langle
q_C(t)+q_1(t)\rangle_{\rm max}}{e} = \beta \frac{E_J}{4E_C},
\end{equation}
where $E_C = e^2/2C_{qb}$ is the charging energy of the qubit and
$\kappa=C_m/C_{osc}\ll 1$. For $\beta < 1$, we can neglect
Landau-Zener transitions and the qubit will follow the oscillator
dynamics adiabatically.

At the charge degeneracy point, we rotate the full Hamiltonian to
the qubit eigenbasis ($\sigma_x \leftrightarrow \sigma_z$) and
expand the eigenenergies to second order in the induced charge
fluctuations
\begin{eqnarray}
\label{TheHamiltonian} H &=& -\frac{E_J}{2}\sigma_z+ \left(
\frac{1}{2C_{osc}} - \frac{g_C}{2}\sigma_z \right) (q_C+q_1)^2 + \frac{\phi_C^2}{2L} +  \nonumber \\
&+& \frac{q_1^2}{2C_c} + \frac{1}{\Delta x}\sum_{i = 1}^\infty
\left( \frac{(q_{i+1})^2}{2C_T} + \frac{(\phi_{i+1} -
  \phi_i)^2}{2L_T}  \right)
\end{eqnarray}
where $g_C=8\kappa^2E_C^2 / e^2 E_J$ indicates the
qubit-oscillator coupling. From the qubit perspective, the slow
transverse charge fluctuations induced by the oscillator result in
second order longitudinal fluctuations as described by the
$(q_C+q_1)^2 \sigma_z$-term\cite{MakhlinOptimalPRL}. The same term
causes a state dependent shift in the oscillator's electrostatic
energy. This may be modelled as an effective oscillator
capacitance $C^{g/e}_{osc} = C_{osc} \pm g_C C_{osc}^2 \equiv
C_{osc} \pm C_Q$, where the last term is the state dependent
\emph{quantum capacitance} $C_Q$. This in turn gives different
oscillator resonance frequencies for the qubit in the
ground/excited state $\omega_0^{g/e} = \omega_0 (1\mp
C_Q/2(C_{osc}+C_c))$, which was recently experimentally measured
in Refs.~\cite{DutyPRL, Sillanpaa}.
%%%%%%%%%%%%%%%%%%%%%%%%%%%%%%%%%%%%%%%%%%%%%%%%%%

{ The same type of
  analysis can be done for the flux qubit in
  Fig. \ref{CircuitFig}\cite{JohanssonJPCM06}. The oscillator part of this Hamiltonian is}
\begin{eqnarray}
H_{\mathrm{flux}}^{osc} = %-\frac{\Delta}{2}\sigma_z +
\frac{q_C^2}{2C} + \left( \frac{1}{2L} -
\frac{g_L}{2}\sigma_z\right)\phi_C^2
%&+&\frac{q_1^2}{2C_c} +
%\frac{1}{\Delta x}\sum_{i = 1}^\infty
%\left( \frac{(q_{i+1})^2}{2C_T} + \frac{(\phi_{i+1} -
%  \phi_i)^2}{2L_T}  \right).
\end{eqnarray}
where now the qubit-oscillator coupling is $g_L=2
M^2\langle\phi_J\rangle^2/\Delta L_q^2 L^2$. Here
$\langle\phi_J\rangle$ is the average flux generated by the
circulating current in the Josephson loop, $M$ is the mutual
inductance between the oscillator and qubit, $L_q$ is the qubit
inductance and $\Delta$ is the qubit level splitting. In this
environment, the
  Josephson junction acts as an effective inductance
$ L_{osc}^{g/e} = L \pm L^2 g_L$ giving the state dependent
resonance frequencies $\omega_0^{g/e} = \omega_0(1\mp L g_L/2)$.
%{ In both cases, the effect of the qubit is the same: shifting the oscillator resonance frequency.}
%describe the quantum noise properties of the driving source and
%transmission line used for detecting this frequency shift.
%%%%%%%%%%%%%%%%%%%%%%%%%%%%%%%%%%%%%%%%%%%%%%%%%%%%%%%%%%%%%%%

%\subsection{Transmission line modes}
The equations for the transmission line coordinates and momenta,
in the limit $\Delta x \rightarrow 0$, correspond to the massless
Klein-Gordon equation, which has traveling wave solutions,
$\phi(x,t) = \phi_{in}(t-x/v) +\phi_{out}(t+x/v)$, with components
moving towards and away from the oscillator  with a velocity
$v=1/\sqrt{L_T C_T}$. The driving source determines the in-field,
$\phi_{in}(t-x/v)$, from which we may derive the detectable
out-field, $\phi_{out}(t+x/v)$, as well as the charge field
$q(t)=q_C(t)+q_1(t)$ and phase field $\phi_C(t)$, giving the
backaction on the charge and flux qubit, respectively. In so
doing, we solve the Heisenberg equations of motion corresponding
to the Hamiltonian in Eq.~(\ref{TheHamiltonian}). Since they are
linear, this is straightforward in the Fourier representation. The
linearity is a consequence of the weak coupling to the qubit,
which is the only non-linear circuit element. However, we are free
to consider any strength of the oscillator-transmission
line-coupling.
%Suppressing the qubit dependence of $C_\Sigma^{g/e}$ for brevity the

First analyzing the charge qubit read-out we define the
state-dependent transconductance
$\chi^{g/e}(\omega)=q^{g/e}(\omega)/\phi_{in}(\omega)$, and write
the general solution
\begin{eqnarray}
\label{chiEq} \phi^{g/e}_{out}(\omega)  &=&
\frac{\chi^{g/e}(\omega)}{\left(\chi^{g/e}(\omega)\right)^\ast}\phi_{in}(\omega)
=e^{i\varphi^{g/e}_r(\omega)}\phi_{in}(\omega), \\
\chi^{g/e}(\omega) & = & \frac{i2C_c C^{g/e}_{osc} L \omega^3} {1-
 (C_{osc}^{g/e}+C_c )L\omega^2-i\omega C_cZ_0(1-L\omega^2C_{osc}^{g/e})}
 , \nonumber
\end{eqnarray}
where $Z_0 = \sqrt{L_T/C_T}$ is the characteristic impedance of
the transmission line. Since there is no dissipation in the lumped
circuit, we have $|\phi_{g/e}^{out}(\omega)|=|\phi^{in}(\omega)|$,
and the reflected signal is characterized by a frequency
dependent phase shift $\varphi^{g/e}_r(\omega)$.
%We should also remember that $q^{g/e}(\omega)$ and $\varphi^{g/e}_r(\omega)$
%depend on the qubit state through $C_\Sigma^{g/e}$.
By introducing creation and
annihilation operators for the in-field satisfying the commutation
relations $[a_\omega,a^\dagger_{\omega'}] =
\delta(\omega-\omega')$ and $[a_\omega,a_{\omega'}] = 0$, we can
write
\begin{eqnarray}
\phi_{in}(x,t) &=& N \int_0^\infty \frac{d\omega}{\sqrt{\omega}}
\left( a_\omega e^{-i\omega(t-x/v)} + h.c. \right), \nonumber \\
\phi_{out}^{g/e}(x,t) &=& N \int_0^\infty
\frac{d\omega}{\sqrt{\omega}}
\left( a_\omega
e^{i\left[\varphi^{g/e}_r(\omega)-\omega(t-x/v)\right]} +
h.c. \right), \nonumber \\
q^{g/e}(t) &=& N \int_0^\infty \frac{d\omega}{\sqrt{\omega}}
\left( \chi^{g/e}(\omega)a_\omega e^{-i\omega t}+ h.c. \right) ,
\end{eqnarray}
with normalization $N=\sqrt{\hbar Z_0/4\pi}$.
%\subsection{Signal detection and Quantum Efficiency}
We model a signal generator by setting the in-field to a coherent
state with a narrow-band distribution around the drive frequency
$\omega_d$ and an intensity corresponding to $\Gamma_{in}$ photons
per second.  Using the
standard model of homodyne detection, the time $t_{ms}$ needed to
distinguish the phase
difference $\delta\varphi_r = \varphi_r^e(\omega_d) -
\varphi_r^g(\omega_d)$ in the out-field is
$t^{-1}_{ms} = 4 \Gamma_{in} \sin^2\left(\delta\varphi_r/2
\right)$ \cite{ZollerGardiner}. We can then compare this to the
measurement-induced qubit
dephasing.  In the weak qubit-oscillator coupling regime,
we straightforwardly calculate the decay of the off-diagonal
element of the reduced qubit density matrix $|\langle \sigma^+(t)
\rangle|=|\langle \sigma^+(0) \rangle|e^{-t/t_\varphi}$ and find the
dephasing rate
\begin{equation}
\label{DephaseRateEq} t^{-1}_\varphi=\Gamma_{in} \frac{(g_C
Z_0)^2}{8 \omega_d^2}
\left(|\chi^g(\omega_d)|^2+|\chi^e(\omega_d)|^2\right)^2 .
\end{equation}
The quantum efficiency $\eta$ can now be evaluated for arbitrary
circuit parameters.  First we note that the drive strength
$\Gamma_{in}$ cancels in the expression for $\eta$. The
LC-oscillator has a quality factor $Q=(C_{osc}+C_c)/C_c^2 Z_0
\omega_0$, and close to resonance we can approximate
Eq.~(\ref{chiEq}) as
\begin{equation}
\label{ChiAppEq} \chi^{g/e}(\omega)=-2\frac{C_{osc}}{C_c Z_0}
\left[
1+i2Q\left(\omega-\omega^{g/e}_0\right)/\omega_0\right]^{-1} ,
\end{equation}
Driving the circuit at the average resonance frequency
$\omega_d=(\omega_0^g+\omega_0^e)/2$ the quantum efficiency is
\begin{equation}
\eta=\frac{t_\varphi}{2t_{ms}}= \frac{\left(x^{-1}+x\right)^2}{8}
2\sin^2{\left[2\arctan{x}\right]} = 1,
\end{equation}
where $x=Q C_Q/(C_{osc}+C_c)$. Somewhat surprisingly, we find a
quantum efficiency of unity independent of $Q$ and the coupling to
the qubit $C_Q/C_{osc}$. For non-optimal drive frequencies the
efficiency is below one. The inefficiency is related to the
information stored in the state-dependent delay time of the
measurement pulse.
%In theory, it is possible to get quantum
%limited read-out even without the LC-circuit, connecting the qubit
%directly to the transmission line. However, the resulting phase
%difference is too small to be detectable in practice.
%I moved this discussion from the end of the paper

For flux qubit read-out Eq.~(\ref{DephaseRateEq}) apply with the
substitution $g_C \rightarrow g_L$ and using the relevant transfer
function
$\chi(\omega)^{g/e}=\phi^{g/e}_C(\omega)/\phi^{in}(\omega)$.
%For the resonance the quality factor has a similar expression as for
%the charge qubit, and the value at resonance is $\chi_{\rm res}=-2i C_c Q/C_{osc}$.
In this case we also find full quantum
efficiency independent of Q.

We can now relate this result to the quantum efficiency of some
other read-out methods.
%The single-electron transistor (SET) biased where transport is through
%sequential tunnelling has a low efficiency $\eta \ll 1$
%\cite{MakhlinMeasurementPRL}, while the superconducting SET biased
%at a special point where transport is based on resonant
%Cooper-pair tunnelling has an efficiency approaching
%one\cite{ClerkSSETPRL}.
Detecting the phase of the signal transmitted through the CPW
cavity in the Yale experiment has half the optimal efficiency
$\eta=1/2$\cite{YalePRA}, where the inefficiency is related to the
reflected signal, which is not detected. In our system, the
efficiency is optimal since the whole signal is reflected and
detected. In the point contact detector, electrons in a range of
energies, as specified by the driving voltage, are used to probe
the contact. Thus a particular energy-dependence of the point
contact transmission is needed for the efficiency to approach
unity\cite{PointContactPRLs}. In our system, the signal source is
essentially monochromatic, probing the system at a single energy.
Thus, there are no constraints on the energy dependence of the
scattering, allowing, e.g. for a low-$Q$ oscillator.

%\subsection{Thermal dephasing in the off-state}
Finding full quantum efficiency for a wide range of parameters
leads us to consider optimizing the circuit parameters for other
criteria. One important figure of merit is the rate of qubit
dephasing induced by the measurement device when {\em not}
measuring. In this case, we assume a thermal distribution with
temperature $T$ of the in-field and find the dephasing time
\begin{equation}
(t^{\rm off}_\varphi)^{-1}=n(\omega_0) \left[1+n(\omega_0)\right]
Q \frac{C_Q^2}{C_{osc}^2}\omega_0,
\end{equation}
where $n(\omega_0)=1/(e^{\hbar\omega_0/k_B T}-1)$ is the thermal
photon occupation number at the oscillator frequency. This
expression is similar to what is found for a superconducting flux
qubit coupled to a DC SQUID\cite{DelftFluxPhotonDephasing}. To
minimize off-state dephasing one should cool the oscillator, but
it is also advantageous to reduce the oscillator-qubit coupling
and to use a low-$Q$ oscillator.
%A low-$Q$ oscillator reduces dephasing by suppressing photon fluctuations at low frequency, moving %them to higher frequency where they do not contribute.
Basically, the integral of the spectral density of the
fluctuations, the variance, is fixed by thermodynamics,
independent of $Q$.  Therefore, lowering $Q$ stretches the same
power over a wider bandwidth, reducing the low-frequency power.

%\subsection{Qubit Relaxation rate}
Another figure of merit is the relaxation time of the qubit,
$t_1$, induced by the measurement device.
%While dephasing at the
%degeneracy point is of second order in the oscillator-qubit
%coupling,
%Because relaxation has a contribution already at first order,
We evaluate $t_1$ using standard weak coupling
expressions\cite{WendinShumeiko2005}, giving that it is
proportional to the real part of the impedance seen from the
qubit, at the qubit frequency $\omega_{qb}$. For the simple
oscillator circuit shown in Fig.~\ref{CircuitFig} with a low $Q$,
$t_1$ could be as short as a few hundred nanoseconds. The solution
to this apparent problem is to add a non-dissipative low-pass
filter between the transmission line and the oscillator. Due to
the large frequency difference between the oscillator and qubit,
this is straightforward. In principle, commercially available pi-filter can improve $t_1$ by a factor of 1000 while
not affecting the low frequency measurement properties. This is
well beyond the point where other sources of relaxation will
dominate.
%\subsection{Optimization for realistic circuit parameters}

Along with these intrinsic figures of merit, we should also
consider the performance of the read-out constrained by existing
technology. In particular, we want to evaluate the measurement
time given that we first amplify the reflected signal with an
amplifier characterized by an impedance, $Z_0$, and a noise
temperature, $T_N$, which exceeds the quantum limit.  We also
assume the input of the amplifier is coupled to a matched load,
e.g. using an isolator. Now, the maximum number of photons in the
oscillator, $n_{max} = C_{osc} (e \beta E_J/2 E_C
C_m)^2/\hbar\omega_d$, is determined by how hard we can drive the
qubit, which is limited by the width of the quantum capacitance
peak in gate charge. Roughly, we can use the half width at half
maximum (HWHM) of the peak, giving $\beta=\sqrt{2^{2/3}-1}/2$ in
Eq.~(\ref{drive_strength_limit}).  Taking the phase shift between
the ground and excited state as $\varphi_r = \arctan(4 Q
C_Q/C_{osc})$, we can estimate the measurement time as
\begin{equation}
t_{ms}=8 \frac{n_{amp}}{n_{max}\varphi_r^2}\frac{Q}{\omega_d} \approx \frac{Z_0 k_B T_N C_c^2}{4 \beta^2\lambda^2 e^2}
\label{RealisticTmsEq}
\end{equation}
where $n_{amp} = k_B T_N/\hbar\omega_d$ is the number of noise
photons in the amplifier, $\lambda = C_m/C_{qb}$, and we have
assumed a signal-to-noise ratio in phase of 2, which corresponds
to the same definition used for the quantum limit.  In the last
approximation, we assume $\varphi_r \approx 4 Q C_Q/C_{osc}$ .

We then want to find a circuit design that gives the fastest
measurement time while still protecting the qubit from thermal
dephasing. In addition, we require that the measurement time be
longer than the response time of the resonant circuit, $\tau_Q =
Q/\omega_0$.  This ensures that the reflected measurement pulse is
not distorted and that its delay is not state dependent.  In Fig.
2, we plot optimized values for $n_{max}$, $\varphi_r$, $Q$, and
$t_\phi^{\rm off}$ as a function of $t_{ms}$. We assume
$t_{ms}/\tau_Q = 4$, $T= 20$ mK, $\lambda= 0.5$, $\omega_0/2\pi =
650$ MHz, $\omega_{qb}/2\pi$ = 5 GHz, $Z_0 = 50 \Omega$  and $T_N
= 0.8$ K \cite{Quinstar} which implies $n_{amp} \approx 25$.  We
see that measurement times of order $50$ ns are readily achievable
while still maintaining off-state dephasing times greater than 10
$\mu$s.
\begin{figure}
\includegraphics[width=6cm]{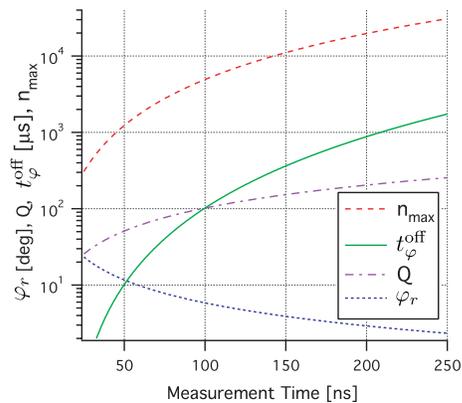}
\caption{The number of oscillator photons ($n_{max}$), off-state
thermal dephasing time ($t_\phi^{\rm off}$), oscillator quality
factor ($Q$), and phase shift ($\varphi_r$) as a function of
read-out time.} \label{OptFig}
\end{figure}

We can compare this result with the estimates of \cite{YalePRA},
trying to understand how the estimates here are 2-3 orders of
magnitude faster.   Fundamentally, the speed-up comes from having
a much lower $Q$, which determines the response time of the
resonator.  However, lowering $Q$ alone also decreases
$\varphi_r$.  In fact, if nothing is done to compensate this, the
measurement time actually increases as $1/Q$.  The way to
compensate is to increase $n_{max}$.  This is where the large
detuning of the our proposal is key.  In the near-resonant design
of \cite{YalePRA}, $n_{max}$ is limited by the fact that the shift
of the qubit frequency due to photon occupation must be less than
the detuning.  If this condition is violated, qubit relaxation is
strongly enhanced by the cavity.  This implies $n_{max} \approx
100$ and $n_{amp}/n_{max}\varphi_r^2 \approx 1$.  This is not a
limitation for a low-frequency oscillator, however, allowing for
much harder driving.  For the the case of $t_{ms} = 50$ ns above,
we have in fact $n_{osc} \approx 1200$, maintaining
$n_{amp}/n_{max}\varphi_r^2 \approx 1/2$ despite the small phase
shift.  In fact, this ratio is constant for the whole range of
plotted parameters.
%In \cite{YalePRA}, the phase shift between states is saturated and $n_{max}\varphi_r^2 \rightarrow n_{max}$.
%\subsection{Summary and Discussion}

In summary, we have discussed how to optimize a general dispersive
qubit read-out for speed, while maintaining maximum quantum
efficiency and long dephasing times.  We have found that reading
out a charge qubit by measuring its quantum capacitance, through
the frequency shift of a lumped circuit LC-oscillator, is quantum
efficient ($\eta=1$) independent of the oscillator quality factor.
%I've moved this discussion closer to the initial result about the oscillator Q
%
%In this approach,
%the whole signal is reflected and detected, which explains the
%doubled efficiency compared to the cavity transmission read-out. In
%the point contact detector the energy range of probing electrons
%is set by the driving voltage, giving constraints on the energy
%dependence of the scattering properties. In our system the
%monochromatic signal source probes the system at a single energy,
%thus relaxing any constraints on the energy dependence of the
%scattering, allowing e.g. for a low $Q$ oscillator.
We also find that a low $Q$ protects the qubit from thermal
dephasing in the off-state, while increasing the speed of the
read-out.  Furthermore, we have discussed how to optimize the
circuit parameters to achieve single-shot read-out using a
commercial amplifier.
%A similar analysis can be made for an
%inductively coupled flux qubit\cite{JohanssonJPCM06}.

We thank Per Delsing, Tim Duty, Vitaly Shumeiko and Margareta
Wallquist for useful discussions. This work was supported by the
Swedish SSF and VR, by the Wallenberg foundation, and by the EU
under the IST-SQUBIT-2 and RSFQUBIT programmes.


\begin{thebibliography}{99}
\bibitem{Quinstar}
Quinstar U-650-2 measured at 4.2K in our laboratory.
\end{thebibliography}
\end{document}